\documentclass{article}
\usepackage[utf8]{inputenc}
\usepackage{authblk}
\usepackage{setspace}
\usepackage[margin=1.25in]{geometry}
\usepackage{graphicx}
\graphicspath{ {./figures/} }
\usepackage{subcaption}
\usepackage{amsmath}
\usepackage{lineno}
\usepackage{comment}
\usepackage{float}
\usepackage{textcomp}
\usepackage[usenames,dvipsnames]{color}
\usepackage[square,numbers]{natbib}
\usepackage{hyperref}
\DeclareUnicodeCharacter{2212}{-}
\usepackage{afterpage}

\title{\bf Layered Multiple Scattering Approach to Hard X-ray Photoelectron Diffraction: \protect\\ Theory and Application}

\author[1,2]{Trung-Phuc Vo}
\author[3,4]{Olena Tkach}
\author[5]{Sylvain Tricot}
\author[5]{Didier S\'{e}billeau}
\author[6]{J\"urgen Braun}
\author[1]{Aki Pulkkinen}
\author[7]{Aimo Winkelmann}
\author[3]{Olena Fedchenko}
\author[3,8]{Yaryna Lytvynenko}
\author[3]{Dmitry Vasilyev}
\author[3]{Hans-Joachim Elmers}
\author[3]{Gerd Sch\"onhense}
\author[1*]{J\'{a}n Min\'{a}r}

\affil[1]{New Technologies-Research Centre, University of West Bohemia, 30100 Pilsen, Czech Republic.}
\affil[2]{Institute of Physics, Czech Academy of Sciences, Cukrovarnická 10, 162 00 Praha 6, Czech Republic.}
\affil[3]{Johannes Gutenberg-Universit\"at, Institut für Physik, 55128 Mainz, Germany.}
\affil[4]{Sumy State University, Rymskogo-Korsakova 2, 40007 Sumy, Ukraine.}
\affil[5]{Univ Rennes, CNRS, IPR (Institut de Physique de Rennes) - UMR 6251, F-35000, Rennes, France.}
\affil[6]{Department Chemie, Ludwig-Maximilians-Universität München, Butenandtstr. 5-11, 81377 München, Germany}
\affil[7]{AGH University of Krakow, Academic Centre for Materials and Nanotechnology, Krak\'{o}w, Poland.}
\affil[8]{Institute of Magnetism of the NAS of Ukraine and MES of Ukraine, 03142 Kyiv, Ukraine.}
\affil[*]{Address correspondence to: jminar@ntc.zcu.cz}
\date{\today}

\begin{document}

\maketitle
\begin{abstract}
Photoelectron diffraction (PED) is a powerful technique for resolving surface structures with sub-angstrom precision. At high photon energies, angle-resolved photoemission spectroscopy (ARPES) reveals PED effects, often challenged by small cross-sections, momentum transfer, and phonon scattering. X-ray PED (XPD) is not only an advantageous approach but also exhibits unexpected effects. We present a PED implementation for the spin-polarized relativistic Korringa-Kohn-Rostoker (SPRKKR) package to disentangle them, employing multiple scattering theory and a one-step photoemission model. Unlike conventional real-space approaches, our method uses a k-space formulation via the layer-KKR method, offering efficient and accurate calculations across a wide energy range (20–8000 eV) without angular momentum or cluster size convergence issues. Additionally, the alloy analogy model enables simulations of finite-temperature XPD and effects in soft/hard X-ray ARPES. Applications include modeling circular dichroism in angular distributions (CDAD) in core-level photoemission of Si(100) 2p and Ge(100) 3p, excited by 6000 eV photons with circular polarization.
\end{abstract}

\section{Introduction}
As an influential technique for probing the atomic structure in the vicinity of a given emitter, photoelectron diffraction (PED) is widely used for various investigation purposes: crystal structures, bonding geometries of atoms, and the local environment of impurity or dopant atoms inside surfaces \cite{fadley1994photoelectron,woodruff2010surface,westphal2003study,greif2013photoelectron}. It is considered analogous to angle-resolved photoemission spectroscopy (ARPES) at the fundamental level, specifically regarding the angular distribution of photoelectrons emitted from a crystal surface. However, the underlying physics and the investigative objectives of the two approaches are different. The angular distribution of the emitted electrons represents the momentum of the initial states in ARPES, while in PED it reveals the interference of photoelectron waves from final states. The observed change in PED intensities depends on the constructive or destructive interference of all the coherently emitted electrons. This means that details related to atomic arrangements, bonding orientations and distances, and electronic and chemical properties are analyzed. Depending on the utilized photon energies with respect to photoelectron kinetic energies, this tool can be termed either ultraviolet-PED (UPD) or X-ray-PED (XPD). In summary, photoemission can provide information about core-level electrons or delocalized valence band electrons. Accordingly, there are core-level PED (CL-PED) and valence-band PED (VB-PED). The main difference is that the angular distribution in VB-PED is caused by two interference phenomena: from primary waves emitted at different atomic sites and from the scattering of corresponding secondary waves at neighboring sites \cite{kruger2018photoelectron}. At high photon energies, XPD effects \cite{gray2011probing,gray2012bulk,kalha2021hard,babenkov2019high,schonhense2020momentum,arab2019electronic,gray2013momentum} obscure ARPES data, along with other complications such as poor cross-sections, considerable momentum transfer from photons causing a pronounced modulation of ARPES patterns, and substantial phonon scattering.

Given the significant contributions of PED (especially in the circular dichroism in the angular distribution (CDAD) interpretation and disentangling of XPD effects), the development of a suitable theoretical model of scattering is of paramount importance. Powerful computational approaches that address single and multiple scattering in real space and reciprocal space have been developed earlier and will be summarized and reviewed in the section \ref{review}. However, these packages share some limitations. First, most of them perform calculations in real space \cite{barton1985small,kaduwela1991application,matsushita2010photoelectron}, which might not be convenient for high kinetic energies where the effective cluster sizes need to be very large. Second, some models have to borrow the atomic potential from other sources. For instance, in the \underline{MSCD} code \cite{chen1997mscd}, it is taken from the muffin-tin potential in the database of Moruzzi \textit{et al.} \cite{moruzzi2013calculated}, which neglects relativistic effects. For low kinetic energy, \underline{MsSpec} \cite{sebilleau2011msspec} utilizes the potential file from Munich SPRKKR \cite{ebert2011calculating,ebert2022munich}. 
In the case of the dynamical high-energy electron diffraction approach to XPD \cite{winkelmann2004simulation,winkelmann2008high}, the real and imaginary parts of the crystal potential (assumed to have 3D bulk periodicity) are computed based on the Fourier coefficients of atomic potentials available in various parameterizations \cite{kirkland2020}.
Third, some approaches are designed to cope with a specific regime of kinetic energy, not flexibly covering the whole range from UV to hard X-ray. For instance, \underline{EDAC} \cite{de2001multiple} is designed to describe diffraction at lower kinetic energies in comparison to the method proposed by Winkelmann \textit{et al.} \cite{winkelmann2004simulation,winkelmann2008high}. Fourth, although hard X-ray calculations (e.g. a chromium K$_\alpha$ source) \cite{williams2012observation,yang2013investigation} attempt to simulate XPD patterns, not all of them are available for CDAD patterns for core levels in this regime, to the best of our knowledge. Indeed, the \underline{TMSP} code \cite{matsushita2010photoelectron} developed by Matsushita and colleagues can work with circularly polarized light \cite{matsui2016photoelectron} and interpret the CDAD in terms of the rotational shift of forward focusing peaks around the incident-light axis \cite{matsui2012photoelectron,maejima2014site}. However, no simulated CDAD for hard X-ray core levels has been published. Fifth, Kikuchi diffraction \cite{kikuchi1928diffraction} is well-documented in scanning and transmission electron microscopy (SEM and TEM) with energy sources up to 100 keV, where it manifests as electron backscatter diffraction patterns \cite{schwartz2009electron}. Nevertheless, it is challenging to observe this phenomenon in photoemission spectroscopy, and few simulations are able to reconstruct it \cite{fedchenko2019high,tkach2023circular,tkach2024asymmetric}. Lastly, to accurately represent scattering by cluster approaches, there is a high need for large values for the maximum angular momentum $l_{max}$ (the number of scattering phase shifts). In the case of 10 keV electrons scattering from an atomic potential with a radius of 1 {\AA} radius, this parameter must be increased to 100 \cite{winkelmann2008high}.

In our earlier effort \cite{tkach2023circular}, major fine features of W 3d\textsubscript{5/2} at $h\nu$=6 keV were relatively well reproduced by calculations. Based on this work and previous research \cite{vo2024analyzing}, which highlighted a more than 50\% contrast in the spin-orbit doublets and the discriminant diffractogram of sub-levels in the case of W(110), this study describes our method utilized in the above-mentioned works in details and resolves these combined obstacles for Si(100) and Ge(100) at $h\nu$=6 keV. Our focus will be on studying fine Kikuchi lines and CDAD for core-level emission XPD (CL-XPD), which is more widely utilized compared to its counterpart, valence-band XPD (VB-XPD) \cite{stuck1995accuracy,herman1992valence,osterwalder1990x,cox1991core}.

\section{Results}
\subsection{Convergence tests} \label{convergence}
This study makes use of the layer Korringa-Kohn-Rostoker (LKKR) scheme in the spin-polarized relativistic Korringa-Kohn-Rostoker (SPRKKR) package, which is based on multiple scattering theory. The theory posits that a system consists of atomic layers, with the scattering characteristics of each layer being computed using a partial-wave basis set. Here, Green's functions are expressed in terms of spherical harmonics and radial functions, allowing the problem to be separated into angular and radial parts. This is where the maximum angular momentum $l_{max}$ comes into play. The layers are then connected in a plane-wave basis to create a solid. The inter-layer scattering is handled by expanding Green's functions into the infinite set of reciprocal lattice vector $\vec{G}_{hkl}$. At this stage, it is the role played by the number of $\vec{G}_{hkl}$. Nevertheless, calculations are carried out by the finite values of these two basis sets. It is not straightforward to predict their proper values as the situation relies on the experimental geometry, photon and final-state energy, and atom types. As a result, the convergent process must proceed with trial tests.

In the beginning, systematic calculations are conducted with varying numbers of $\vec{G}_{hkl}$ vectors and final state partial waves (orbital angular momenta $l_{max}$). Hundreds of separate simulations using various parameter sets lead to the exploration of significant elements influencing the fine structure of CDAD. As seen in Fig.~\ref{fig:Fig3-1-1.jpg}a, increasing the number of $\vec{G}_{hkl}$ brings more valleys, which become more sharp, besides changing the intensity. The calculations within 49-89 $\vec{G}_{hkl}$ are in poor agreement with the rest. It starts to converge from 137 $\vec{G}_{hkl}$ due to the common tendency. More Kikuchi lines are introduced as a function of the number of $\vec{G}_{hkl}$ in Fig.~\ref{fig:Fig3-1-3.jpg}. On the other hand, under $l_{max}$ variation, the position of intensity peaks and valleys relatively remain the same with respect to $\theta$ angles in Fig.~\ref{fig:Fig3-1-1.jpg}b. Taking a closer look, the shape of these peaks and valleys as well as their vicinity are smoothened, leading to smearing out diffraction patterns in Fig.~\ref{fig:Fig3-1-4.jpg}. Convergence tests are done up to $l_{max}$ = 11 and we observe that qualitatively the position of the photoemission peaks as well as their positions do not change for $l_{max}>$ 3, except the proximity of $\phi$=4\textdegree. In an attempt to balance between accuracy and efficiency, $l_{max}$ has to be truncated at 4 based on the basis of computational time and memory limitations.

Fig.~\ref{fig:Fig3-1-3.jpg} shows a series of total-intensity calculations I\textsubscript{TOT} = I\textsubscript{RCP} + I\textsubscript{LCP} (which possess fingerprints of Kikuchi bands as mentioned in the work of Fedchenko and co-workers  \cite{fedchenko2022structure,fedchenko2019high,fedchenko2020emitter}) for the Si 2p\textsubscript{3/2} (first row) and Ge 3p\textsubscript{3/2} (second row) core levels at $h\nu=6000$ eV. One crucial impression is obviously visible when enhancing the number of $\vec{G}_{hkl}$ such that the diffraction patterns get increasingly complicated since each family of lattice planes creates a new "Umklapp channel" in the diffractogram. The fine structure of the pattern grows more sophisticated in the series 25, 97, and 185 $\vec{G}_{hkl}$-vectors, where Kikuchi patterns vary on a very tiny k-scale. The Kikuchi diffraction in PED is distinct from conventional electron diffraction in that the emitter atom inside the material serves as the source of the diffracted electron wave. The scattered wave is diffracted at the crystal's lattice planes before the scattered electrons reach the detector on the vacuum side. The Kikuchi diffraction mechanism is distinguished by incoherent yet localized electron sources within a crystal. PED enables the differentiation of chemically distinct emitter sites in a crystal due to the element-specific binding energies of the photoelectrons, which is unique to the one in SEM and TEM. The photoelectron intensities are observed as a fine structure in their angular distributions when the measurements were conducted with high enough angular resolution  \cite{fedchenko2019high}. In the experiment, these filigree structures have not been detected, probably due to the restricted $k$-resolution. Nevertheless, there is another issue to blame for this. By reducing Bragg angles, the distances of the electron trajectories in real space rise (see Fig. 3 in Ref.  \cite{tkach2023circular}). The scattering cross section is big, especially for high-Z materials such as tungsten, and the Kikuchi bands get more and more suppressed when indices increase. Conversely, there are a finite number of contributing $\vec{G}_{hkl}$ in the experiment.

Methodical computations of Si 2p\textsubscript{3/2} (a-c) and Ge 3p\textsubscript{3/2} (d-f) in accordance with the diverse value of $l_{max}$ are performed in terms of total intensity patterns I\textsubscript{TOT} in Fig.~\ref{fig:Fig3-1-4.jpg}. The results are obtained from $h\nu=6000$ and 37 $\vec{G}_{hkl}$-vectors. In general, the main Kikuchi bands show some reduction in intensity and spread out more both horizontally and vertically. As $l_{max}$ values change, the diffraction patterns from the background become more spread out. Furthermore, the contrast gradually increases, particularly for the overall intensity, making the black Kikuchi bands and bright center stand out more against the background. $l_{max}$ affects the intensity of the background diffract as typically seen in the high-to-low magnitude variation from $l_{max}=3$ to $l_{max}=7$ (Fig.~\ref{fig:Fig3-1-4.jpg}(d-f)). This influence also happens for Si but not with displayed values of $l_{max}$. In cluster approaches, this convergence parameter is very high (e.g. 100 \cite{winkelmann2008high}) and must be handled with care as the computation time required to calculate the scattered wave function is directly proportional to $nN^{2}(l_{max}+1)^{3}$, where $n$ represents the scattering order, and $N$ represents the number of atoms utilized in the cluster \cite{despont2006x}. However, in our method, we do not have to suffer from its high-value thanks to a mixed basis set of partial waves and plane waves. When using $l_{max}=4$, it is possible to achieve satisfactory agreement between simulations and experiments in subsection \ref{comparison}.

With the aim of further cross-checking the feasible $l_{max}$ integer, in Fig.~\ref{fig:Fig3-1-2.jpg} we utilize a cluster approach based on multiple-scattering spherical-wave cluster basis sets from the \underline{MsSpec} program  \cite{sebilleau2006multiple,sebilleau2011msspec}. The phase shifts have been calculated by a sophisticated Hedin–Lundqvist exchange and correlation potential  \cite{hedin1970effects,hedin1971explicit,fujikawa2000self} so as to describe the finite electron mean free path in the final state via imaginary parts. The incident as well as emission angles and photon source are determined identically to experiments and the one-step model setup. We use unpolarized light and the non-relativistic mode. Being interested in high-energy PED, the deviations from bulk physics owned by surface atoms are sparse  \cite{gewinner1994x}. As a result, we take into account bulk-terminated surfaces. The radius of our spherical cluster is set to 27.15\textup{~\AA} (Fig.~\ref{fig:Fig3-1-2.jpg}a). Here, we investigate the test of cross-section, an intensity-related quantity, by tuning $l_{max}$ values (4, 16, and 24) in the azimuthal scan of Si 2p$_{3/2}$ at $\theta=5^{\circ}$. Apparently, there is a gap among 3 cross-section behaviors, such as magnitude overall and the shape of the curve (especially in the vicinity of $\phi=30^{\circ}$ and $\phi=60^{\circ}$). The result from $l_{max} =4$ seems a bit far from convergence as compared to the other two. Even so, its peaks and valleys share good relative positions (namely, in the surrounding of $\phi=30^{\circ}$ and $\phi=60^{\circ}$) in common with the rest. More examples are addressed when $\phi$ is in between $50^{\circ}$ and $60^{\circ}$.

\subsection{Final-state energy dependence} \label{finalenergy}
The final-state energy $E_{Final}$ is defined as the kinetic energy of the photoelectrons inside the crystal and the quantity in charge of diffraction dynamics. The expected variation amongst calculated patterns at individual final-state energies [4690,6690] eV is seen in Fig.~\ref{fig:Fig3-2-1.jpg} for Si (top row) and Ge (bottom row), respectively. From the simulation point of view, when energy is raised, a pronounced system of Kikuchi bands dominates the diffraction patterns, forming a rich fine structure. Each band consists of a projection plane sandwiched by two lines and its width is determined by the respective reciprocal lattice vector. For consistency in comparison, the entire calculations are designed by the identical value of momentum range and resolution. Overall, there is an agreement concerning the width of observed bands and the position of lines. Fine structures possess a horizontal and a vertical mirror plane that is parallel to the [010] and [001] directions, respectively. Another general feature is that the distinct central zone indicated by a green dash square, can be noticed easily regardless of final state energies.

Fig.~\ref{fig:Fig3-2-1.jpg} (a-d) shows computed diffractograms of Si 2p\textsubscript{3/2} core-level at the (100) plane with 161 $\vec{G}_{hkl}$ and $l_{max}=4$. Via visual inspection, there are several noticeably rapid alterations besides the above-mentioned common behaviors. Firstly, the texture inside the central zone evolves by enhancing $E_{Final}$. Four various symbols are found in the surroundings of the center zone axis [100], running from left to right: bow-tie dartboard (a), shield (b), 4-pointed star cross (c), and butterfly (d). However, all geometrical patterns still follow 4-fold symmetry, i.e. the dartboard pattern is obtained by rotating a bow-tie by 90\textdegree. Another shape turn can be noticed through the distance decrease of a line pair (marked as a yellow dash with a label (1)). Starting from 4690 eV, it is definite to see the space between them. Then, this gap continuously reduces in between (b-c) and seems to become 0 at 6690 eV. Consequently, some bright spots (e.g. four magnificent bright ones labeled (2) in (b)) become hidden. One more typical instance is the deformation of the bright area (orange dotted curves labeled 3 in (a)) along diagonal directions at four corners. These phenomena are derived from the bound relation between the angular range and Bragg angles. In the context where forward scattering appears overriding, the greater the final-state energy is, the smaller the Bragg angle is. Thus, there are some shifts and re-arrangements of fine Kikuchi lines meanwhile main features (horizontal and vertical bands as well as their intersections) remain unchanged.

Analogously, the diffractogram of Ge 3p\textsubscript{3/2} is analyzed in Fig.~\ref{fig:Fig3-2-1.jpg} (e-h). Because of the mismatch related to photoelectron wavelength (caused by different core-level binding energies), the patterns are markedly dissimilar from those in Fig.~\ref{fig:Fig3-2-1.jpg} (a-d). The number of $\vec{G}_{hkl}$ (113 in this case) is also one of the essential causes of reshaping patterns (as discussed in Section \ref{convergence}). The major Kikuchi grid is recognizable and aligned with the one from Si. For consistency and convenience in qualitatively comparing with Si outcomes, the color bar is fixed. Hence, plots are not in good contrast and it is a bit hard to identify diffraction which is contributed by thin Kikuchi lines. Despite that, a striking difference regarding global intensity values is captured when bright areas are larger in general. As the energy rises, pattern adjustment is forcibly determined by the diffraction angle. As an illustration, the angle between two narrow fine lines which are in yellow dash (e) declines and its vertex position moves up when switching from 4690 eV to 5190 eV. Continuously with the enhancement of energy to 6190 eV, the bright spot tagged by (4) in (f) is fading. In Kikuchi photoelectron diffraction, there exists an orientation of higher and lower intensity. This bright-dark distribution can be interpreted in terms of the reciprocity principle and the coupling probability of photoelectron from localized emitters with outgoing wave  \cite{winkelmann2008high}.

\subsection{Experiments meet theories} \label{comparison}
Fig.~\ref{fig:Fig3-4-1.jpg} displays a comparison of measured and calculated CDAD in pairs for the Si 2p\textsubscript{3/2}. The photoelectron diffraction pattern is simulated over a polar angle range of $0^{\circ} \leq \theta \leq 10^{\circ}$ and over a full $360^{\circ}$ azimuthal range ($\phi$) with 180 points for both angles at the photon energy $h\nu=6$ keV. The pattern centers are indicated by +. Band edges are marked by full arrows on the left-hand side of the figure, and band centers are marked by dotted arrows. The CDAD signal (CDAD = I\textsubscript{RCP} - I\textsubscript{LCP}) is a difference between the intensities of right and left circular-polarized light that emphasizes faint details in the patterns. When this difference is normalized, it leads to a so-called CDAD asymmetry A\textsubscript{CDAD} = (I\textsubscript{RCP}($k_{x},k_{y}$) - I\textsubscript{LCP}($k_{x},k_{y}$))/(I\textsubscript{RCP}($k_{x},k_{y}$) + I\textsubscript{LCP}($k_{x},k_{y}$)) reflecting the symmetry behavior of CDAD. This normalized factor is a good choice for experiment-theory comparisons due to its autonomy of the spectrometer transmission function and the free-atom differential photoelectric cross-section. As anticipated, the A\textsubscript{CDAD} is antisymmetric referring to the horizontal mirror plane. This "up-down antisymmetry" is an explicit consequence of the atomic CDAD's symmetry (see Fig. 1 in Ref. \cite{tkach2023circular}) and the Kikuchi diffraction's characteristics (see Fig. 2 in Ref. \cite{tkach2023circular}). The crystal-lattice mirror plane is positioned horizontally to maintain the antisymmetry. The agreement between observed and computed intensity (a-b), CDAD difference (c,d), and A\textsubscript{CDAD} (e,f) looks quite reasonable. Particularly, for the intensities I\textsubscript{RCP} $+$ I\textsubscript{LCP} the principal Kikuchi bands (as labeled by black arrows) and the 4-pointed star cross clearly show up in Fig.~\ref{fig:Fig3-4-1.jpg}(a-b). Two elliptic shapes (marked as green) are also captured. Compared to the measured total intensity, the contrast is lower and there are more visible minor Kikuchi bands from calculations. The reason for these differences originates from the $l_{max}$ effect (above-mentioned in Fig.~\ref{fig:Fig3-1-4.jpg}(a-c)). The issue is solvable by increasing $l_{max}$ but for the sake of clearly displaying fine structures in detail, the small converged value is opted. It is likely that the CDAD image (Fig.~\ref{fig:Fig3-4-1.jpg}c) shares commonly observed patterns with the total intensity (Fig.~\ref{fig:Fig3-4-1.jpg}a), such as a large diamond shape around the center and two ellipses (marked as green). These similarities are also nicely captured by the one-step photoemission model (Fig.~\ref{fig:Fig3-4-1.jpg}d). The CDAD difference I\textsubscript{RCP} $-$ I\textsubscript{LCP} depicts a higher contrast than the sum of intensities I\textsubscript{RCP} $+$ I\textsubscript{LCP}. However, the 4-pointed star cross, surrounding the CDAD center, is practically hidden in the experimental image (c). From the computational point of view, this footprint is reproduced charmingly in Fig.~\ref{fig:Fig3-4-1.jpg}d. Several band edges are present in the agreement as well, in particular two crossing points (yellow arrows in (c-d)). Likewise, the latter emerges from the A\textsubscript{CDAD} patterns as the vertex of small blue and red triangles, identified by the black arrows in (e,f). There is no doubt that the resolution of A\textsubscript{CDAD} from measurements is not in good quality in comparison with the two above-mentioned ones. Overall, the consensus between the experimented and simulated A\textsubscript{CDAD} is far from perfect. But the mirror plane and some characteristics (namely Kikuchi bands forming a central diamond) still emerge. Calculated spectra have sharp features that need to be broadened to match the momentum resolution of the experiment. To make theoretical calculations more comparable to experimental data, we apply convolutions with a Gaussian function in which the standard deviation $\sigma=2$ (Fig.~\ref{fig:Fig3-4-1.jpg}(\text{b$^{*}$}, \text{d$^{*}$}, \text{f$^{*}$}). For Fig.~\ref{fig:Fig3-4-1.jpg}\text{b$^{*}$}, the Perona-Malik (anisotropic diffusion) filter is employed for I\textsubscript{TOT} to smooth the results while preserving the main spectral features. The "weight" parameter $\lambda$ is set to 1 in this case. In general, convolution steps significantly improve the visual comparison.

To further clarify the parallels between measuring and calculating, we conducted the same measurement as shown in Fig.~\ref{fig:Fig3-4-1.jpg}a but with higher resolution in Fig.~\ref{fig:Fig3-4-1-1.jpg}a. Next, we examine the specifics from a quadrant of A\textsubscript{CDAD} in the second row of Fig.~\ref{fig:Fig3-4-1-1.jpg}. The superior Kikuchi bands are marked with dashed lines in the experimented intensity (a) and A\textsubscript{CDAD} (c). These line patterns are then mapped to the corresponding calculated ones (b,d). The Kikuchi-band edges and crossing points take place in a well-founded arrangement in the matching experiment (a) and simulation (b). Dark (addressed by a dashed arrow) and bright (pointed by a solid arrow) spots located in the corner of the diamond shape are observed in the relevant calculation (Fig.~\ref{fig:Fig3-4-1-1.jpg}b). The computed and measured patterns of the total intensity are arranged in such a way that features can be reflected via a mirror plane. The rough red-blue texture of experimental A\textsubscript{CDAD} is analogous to the theoretical one with respect of not only color but also relative positions.

Next, we turn our attention to the Ge(100) cases in Fig.~\ref{fig:Fig3-4-2.jpg} which can be interpreted by the formalism like in Si(100) diffraction. In Fig.~\ref{fig:Fig3-4-2.jpg}(a, b), the intensity from two light helicities are summed for Ge 2p$_{3/2}$. An inset of a higher-resolution measurement is added to the experimental pattern. The relevant area in the simulation is depicted via a blue dashed circle. Fig.~\ref{fig:Fig3-4-2.jpg}(c, d) shows the difference between these helicities
while Fig.~\ref{fig:Fig3-4-2.jpg}(e, f) belongs the CDAD asymmetry. The "four-fold" geometry marked with blue curves in the intensity difference indicates the matching. The grid made up of Kikuchi lines (dash dots) is superimposed on the diffractogram to conveniently clarify the parallels between measuring and calculating A\textsubscript{CDAD}. Notable features are easy to map between experimental and theoretical interrelationships: vertical (except CDAD patterns, as they are almost invisible) and horizontal Kikuchi bands and the pattern at four corners (indicated by green circles). More specifically, the color flipping of the pattern from upper to lower corners in Fig.~\ref{fig:Fig3-4-2.jpg}(c-f) is triggered by the horizontal mirror plane which is a typical feature between CDAD and A\textsubscript{CDAD}. Besides, black and white spots of the intensity difference (blue and red ones in case of A\textsubscript{CDAD}) formed by several visible band edges are sighted by green arrows. Although recognizing several systematic hallmarks, it seems we are far from reaching good consensus in comparison. From measured outcomes (especially in Fig.~\ref{fig:Fig3-4-2.jpg}c, there is a generic diamond-shaped pattern which is wide with a sharp rim. Nevertheless, it appears quite differentiable in simulation. Another distinguished discrepancy is the distorted version of the bow-shaped feature 1 (yellow dash in Fig.~\ref{fig:Fig3-4-2.jpg}(c,e)). Also, we cannot ignore the fact that in CDAD and A\textsubscript{CDAD} fine diffraction at the upper and lower corners is not in harmony with respect to fine details. To enhance the comparability between theoretical calculations and experimental data, convolutions are applied with a Gaussian function characterized by a standard deviation of $\sigma$ as detailed in Fig.~\ref{fig:Fig3-4-2.jpg}(\text{b$^{*}$}, \text{d$^{*}$}, \text{f$^{*}$}).

Subsequently, we present the diffraction pattern of another core level, Ge 3p$_{3/2}$, as shown in Fig.~\ref{fig:Fig3-4-2-1.jpg}. The similarities between measured and calculated diffractograms are easily found. For example, the Kikuchi lines, characterized by their very low intensity, are detected (denoted by black solid arrows in Fig.~\ref{fig:Fig3-4-2-1.jpg}(a, b)). In addition, the bright fourfold cross (indicated by green dashed arrows) is reproduced though its contrast to the background is not as high as in experiments. The inner diamond shape surrounding the center of this cross also appears in the simulation. The nice agreement is further confirmed by considering the signal subtraction (Fig.~\ref{fig:Fig3-4-2-1.jpg}(c, d)). The faint patterns are likely symmetrized through a vertical mirror plane separating two diffractograms. Obviously, the edges of the big diamond can be recognized via blue dashes. In pursuit of fleshing out the similarities between measured and calculated patterns, it is worth looking at their A\textsubscript{CDAD} in depth (Fig.~\ref{fig:Fig3-4-2-1.jpg}(e, f)). The agreement is not ideal overall. However, the mirror plane and certain features (specifically the formation of the "red-blue" hourglasses inside and outside the big diamond) still become apparent. Taken together, the main findings indicate a greater result of Si 2p$_{3/2}$ and Ge 3p$_{3/2}$ compared to the one of Ge 2p$_{3/2}$. Despite owning the same information in terms of crystal structure and period in the periodic table, it is understandable to experience the difference between their diffraction patterns due to dissimilar electron configurations and investigated core levels. To improve the consistency between theory and experiment, the calculated spectra are convolved with a Gaussian function of standard deviation $\sigma=2$, as illustrated in Fig.~\ref{fig:Fig3-4-2-1.jpg}(\text{b$^{*}$}, \text{d$^{*}$}, \text{f$^{*}$}). This approach enhances the agreement between theory and experiment.

It is distinguished between the instrument resolution and the width of the features in a specific k pattern. The former refers to the intrinsic performance limit of the experimental setup and has been determined to be 0.03 \r{A}$^{-1}$. The latter is the width of the features observed in a specific measurement or image, which is affected by several factors. This width can be much worse than the instrument resolution when the sample quality is not good or the X-ray beam hits a "bad spot" on the sample. After cleaving, it is often complicated to find a good spot. In some cases, there may also be some misalignment, e.g., incorrect sample distance, or the photon beam is not optimally adjusted (e.g., too large a footprint, which also spoils resolution). In many cases, the statistics are not sufficient, so we have to apply a Gaussian blur. This is analogous to the width of a core level signal in comparison with the resolution of an electron spectrometer. Figure 8a and the inset of Figure 9a are shown with better resolution and contrast because the corresponding measurements were done after we improved the setup and conducted the experiment in a smaller k-range. The rest of the figures come from the initial measurements. The instrument resolution is at least as good as the narrowest lines we observe. When lines appear broader, they may arise from several other factors, as discussed above. It is important to note that not the eye-catching broadest features reflect the instrument resolution, but the narrowest lines (which may be rather weak in intensity) in the pattern.

 While the overall agreement between theory and experiment remains qualitatively reasonable, we acknowledge that some quantitative discrepancies persist. From our perspective, several factors may contribute to these differences. First, the choice of effective potential and exchange-correlation functional plays a crucial role. We currently use the local density approximation (LDA) to describe the exchange-correlation potential, which may not be optimal. Previous studies have shown that the choice of potential can strongly influence theoretical predictions. For example, McGovern et al. \cite{mcgovern1979atomic} reported discrepancies arising from a suboptimal Te potential, which were later addressed by Wendin \cite{wendin1981importance} using the random phase approximation with exchange (RPAE) \cite{wendin1976random}, yielding more accurate atomic photoionization cross-sections for Te 4d. Additionally, while the influence of the exchange-correlation functional is often minor in small systems \cite{pueyo2016performance}, its impact becomes more pronounced in complex materials \cite{golze2020accurate}. The LDA is known to overbind, potentially leading to systematic deviations \cite{yuk2024putting,becke2014perspective}. In our calculations, we use the atomic spheres approximation (ASA), which is similar to the muffin-tin approach but ensures full space-filling by allowing overlapping spheres and introducing empty spheres in the interstitial regions. However, ASA imposes shape-related constraints on the potential, and its accuracy can be improved by adopting the full-potential version of the SPR-KKR code \cite{minar2005multiple}. Second, inelastic scattering effects may contribute to the observed differences. Inelastic scattering corrections to the elastic photocurrent (see, e.g., Borstel \cite{borstel1985theoretical} and Braun et al. \cite{braun2018correlation}) are included via a parametrized, complex inner potential. While these effects are considered in the current work, they have not been systematically investigated. A more detailed study of their influence will be presented in a forthcoming publication \cite{vo2025unveiling}. Third, lattice vibrations must also be taken into account. The experimental measurements were conducted at 30 K, while our calculations were performed at 0 K. Although this temperature is low, vibrational damping effects are still relevant. We plan to analyze this temperature dependence more systematically using the Debye-Waller model and the alloy analogy approach within the coherent potential approximation (CPA) in a future publication aimed at reproducing the Si hard X-ray experiments \cite{tkach2024asymmetric}. Finally, crystal imperfections represent another possible source of discrepancy. Our theoretical simulations assume a perfect, defect-free crystal structure. However, real experimental samples may contain hidden structural imperfections such as local strain or disorder, which are known to significantly affect photoelectron diffraction patterns [15,16]. These effects are not captured in our current model but are important to consider in explaining the remaining discrepancies.

\section{Discussion}
The combination of CDAD and XPD techniques provides comprehensive information on the crystal structure and chemical composition of the system being studied. Most of the time, it is rarely accessible to obtain all of this information readily. The challenges in interpreting emitted-electron spectra evolve from the intricacy of the photoemission itself. To extract it, one must make assumptions about a particular physical model and apply the relevant theoretical framework. Even so, a good qualitative analysis demands ab-initio computations and this process may include several levels of complexity. In this study, we concisely outline the interplay and importance of XPD and CDAD. Furthermore, several standard theoretical models for PED are reviewed with an emphasis on physical content and terminology rather than concentrating on mathematical formalism. In the context of high photon energies, we emphasized our fingerprint that can tackle the disadvantages of other codes. XPD diffraction computed in reciprocal space is convenient to compare with momentum microscope measurements. The atomic potential is obtained from our own code, making the XPD results more controllable in a closed simulation loop. The observed outcomes were reproduced in the hard X-ray regime. The upcoming Ge work concerning the full-field photoelectron diffraction and circular dichroism texture will complete the energy range of applications.CDAD signals and asymmetries in the manner of CL-XPD were simulated at 6 keV. The diffraction patterns are significantly altered by the Kikuchi diffraction process. Furthermore, simulating Kikuchi patterns in spectroscopy encourages additional measurements and energizes the current trend in XPD \cite{Fedchenkomagnetic}. The representation of scattering by our approach requires smaller values for the maximum angular momentum than others (e.g. cluster approaches), helping to avoid computationally intensive calculations. A series of convergence parameters (the number of $\vec{G}_{hkl}$ and $l_{max}=4$) was tested. Next, the optimal values were selected for the calculations for Si and Ge core levels, yielding fairly nice harmony with experiments. Fine patterns are well-mapped from the total intensity and CDAD signal. The final-state effect was systematically investigated and had a strong influence on the pattern evolution. Thus, careful selection and control of the kinetic energy are essential for optimizing photoelectron diffraction studies. With this endeavor, we aim to bring another tool for XPD analysis and CDAD interpretation. 

\section{Methods}
\subsection{Brief review of PED simulation progress} \label{review}
Basically, there exist three classifications for adequate PED simulation for solids as presented in Fig.~\ref{fig:Fig2-1.jpg}: cluster-based models, layer-by-layer approaches, and the so-called "lattice-plane" methods.

The first group (Fig.~\ref{fig:Fig2-1.jpg}a) is a right call for short-range probes (e.g. PED or Auger electron diffraction) and broken crystal symmetry (for instance by disorders). Because of the electron's short inelastic scattering length, scattering can be limited to a cluster of individual atoms surrounded by the photo-emitter. A model can be formulated by either single-scattering cluster (SSC) \cite{fadley1979determination,kono1978azimuthal,fadley1984angle,fadley1992synchrotron,chambers1987high,armstrong1985analysis,sinkovic1991theory} or multiple-scattering cluster (MSC) \cite{barton1985small,kaduwela1990assessment,westphal1994photoelectron} methods. For comparison in this subset, the most precise interpretation is MSC with spherical waves \cite{kaduwela1991application}. Specifically, the multiple-scattering effects have to be taken into account while investigating single-crystal because the XPD patterns originate from photoemission of ten (and more) surface layers. Inspired by the work of Kaduwela and co-workers \cite{kaduwela1991application} where multiple scattering (MS) is derived from the Rehr-Albers (R-A) separable propagator approximation \cite{rehr1990scattering} and applied for initial and final states, more CD measurements are nicely reproduced \cite{kaduwela1995circular,van1996application}. Furthermore, the experiment-theory match is improved by utilizing an accurate representation of the Green’s function propagator for evaluating MS expansion \cite{de2001multiple,ynzunza2000circular}. Thereafter, more indispensable modifications and implementations come into existence as seen in following MSC packages: \underline{MSCD} \cite{chen1997mscd}, \underline{PAD2} \cite{harp1998computation}, \underline{MSPHD} \cite{gunnella2000msphd}, \underline{EDAC} \cite{de2001multiple} and \underline{MsSpec} \cite{sebilleau2011msspec}. Recently, Rehr \textit{et al.} \cite{rehr2022real} has discussed the development of real-space Green's function approach for PED based on the R-A formalism and boosted the importance of sharing theoretical software among groups. Workflow tools such as \underline{AiiDA} \cite{huber2020aiida} and \underline{Corvus} \cite{kas2021advanced} were proposed to enable advanced and efficient computation of PED without requiring significant changes to the original code.

Besides, there is another cluster code called \underline{TMSP} \cite{matsushita2010photoelectron,matsushita2020theory}, which relies on photoelectron holography (PEH), a rapidly growing and interconnected field of PED. In \underline{TMSP}, multiple scattering is fully employed, but backscattering is excluded as its intensity appears very weak and negligible when the kinetic energy of the photoelectron is about from 100 eV to several keV. Both PEH and PED can be applied to measure the angular distribution of core-level intensities. Nevertheless, the former is capable of performing three-dimensional (3D) atomic reconstruction. Hence, the direct reconstruction of the local settings of interest around an atom is obtained. In principle, PEH comprises holography and photoelectron diffraction \cite{barton1990photoelectron}. In this manner, the angular distribution of intensities is labeled a hologram that is transformed into a real-space image of the atomic structure. In other words, a hologram is a record of diffraction patterns and interference patterns between two kinds of photoelectron spherical waves. The first is the one directly emitted from the atom (emitter) where photoexcitation occurs. The second is the wave elastically scattered by neighboring atoms (scatterer). These two are the reference and object waves of the optical holography in turn. For further details, there exist many worthy reviews concerning PEH and PED applications \cite{fadley1994photoelectron,chambers1991epitaxial,woodruff1994adsorbate,barton1988photoelectron,harp1990atomic,matsushita2004new} as well as their recent interrelated progress \cite{daimon2018overview,tsutsui2020analyses,kinoshita2019progress,fedchenko2022structure,chasse2018theory,yokoya2022photoelectron,kuznetsov2014x,daimon2020atomic,kuznetsov2018photoelectron}.

On the other hand, the second class (Fig.~\ref{fig:Fig2-1.jpg}b) considers a crystal as a collection of identical layers and makes use of the translational symmetry along surfaces, which is beneficial for describing interlayer multiple scattering in a finite-depth crystal or a surface formed by inequivalent layers on a substrate of repeat periodicity. This category is developed on the basis of low energy electron diffraction (LEED) \cite{pendry1974low,pendry1976theory,li1979binding,li1978multiple}. Recently, the one-step model (OSM) of photoemission \cite{braun2018correlation} is proposed to conduct a nicely theoretical prediction of the core-level problem of W(110) \cite{tkach2023circular} where the final-state wave function is represented by a so-called time-reversed LEED-state. A 50\% difference in the spin-orbit doublets of W(110) was reported in the work of employing circularly polarized radiation. Herein, the sub-levels of 3d and 4p created by SOC splitting (as well as possibly via multiplet splitting and circular polarization) have distinct diffractograms \cite{vo2024analyzing}. This generalized angle-resolved core-level photoemission takes into account the fully relativistic effects and experimental geometry. OSM is applicable not only in reciprocal space but also in real space. Arising from the above-mentioned cluster models, Kr\"uger \textit{et al.} \cite{kruger2011real} suggested a real-space multiple scattering method integrating OSM and claimed good-agreement VB-PED for Cu(111), which is comparable to the result of k-space calculations. Lately, a reasonable comparison has been made between very low energy electron diffraction (VLEED), CL-PED, and VB-PED for monolayer graphene by the variational embedding method including OSM \cite{krasovskii2022one} from the team of Krasovskii. Even though the specifics may be somewhat diverse, diffraction patterns overall display comparable broad elements. Based on LEED theory, it turns out that more hybrid schemes reveal from simple to sophisticated modifications to elucidate complex surface structures. Deriving from tensor LEED (a perturbation technique to high-speed computation of LEED intensity) introduced by  Rous \textit{et al.} \cite{rous1986tensor}, PED intensity is able to be computed with a little adjustment to how photoelectron emitter atomic displacements are handled \cite{omori1999photoelectron}. Combined with Helmholtz's reciprocity theorem, which was initially stated for optics, the path-reversed LEED was effectively employed for PED \cite{pauli2001path}. Next, it was extended to crystal surfaces with many atoms per unit cell, leading to the fact that the outcomes are almost identical to those from traditional forward-path computations \cite{poon2002path}.

When the photoelectron kinetic energies fall into the range of several keV, the hard X-ray photoemission diffraction signal is dominated by scattering in the full, three-dimensional, periodicity of the crystal structure, which results in pronounced bulk features and vanishing sensitivity to surface effects. 
Tailored to this regime, a third simulation group (Fig.~\ref{fig:Fig2-1.jpg}c) is based on the dynamical theory of high energy electron diffraction, adapted to the case of photoelectron sources in a single crystal.
The approach presented in \cite{winkelmann2004simulation} uses the Bloch wave approach to electron diffraction \cite{bethe1928adp}, with a plane wave expansion of the diffracted electron waves in a bulk crystal \cite{spencezuo1992emd}.
The basic building blocks of the theoretical Bloch wave simulation model are the Fourier coefficients of the real and absorptive parts of the 3D crystal potential with corresponding reciprocal lattice vectors, and it can be conceptually advantageous that these theoretical entities are in very direct correspondence to experimentally observed Kikuchi band features \cite{kikuchi1928diffraction}. 
In the context of Kikuchi-band theory, high kinetic energy PED patterns from single crystals can thus be meaningfully discussed in terms of photoelectron reflections at lattice planes, with a Kikuchi band having a width of twice the corresponding Bragg angle \cite{siegbahn1970angular,goldberg1980explanation,trehan1987single}.
Examples of core-level XPD in the hard X-ray range interpreted by Bloch wave simulations can be found in Refs. \cite{winkelmann2008high,fedchenko2019high}.

\subsection{Theoretical model}
The \textit{ab-initio} calculations provided are based on density functional theory with full relativistic effects, using the multiple scattering Korringa-Kohn-Rostoker Green's functions in the SPRKKR package \cite{ebert2011calculating,ebert2022munich}.  In this current research, we have freshly reformulated and generalized angle-resolved core-level photoemission, which includes XPD effects as per the work of Schlath{\"o}lter and Braun \cite{schlathlter1999theorie,plogmann1999local,bansmann1999relationship}, enabling us to cover a wide kinetic-energy range up to the hard X-ray regime. This technique accounts for finite temperature diffuse scattering effects, resulting in I\textsubscript{non\_direct} transitions (Eq. (4)) using the alloy-analogy model \cite{braun2013exploring}. The methodology proposed for computing the final state in XPD differs from the standard real-space dynamical multiple scattering computations \cite{de2001multiple,sebilleau2011msspec}. It utilizes the layer KKR method to depict an infinite stack of layers. Relativistic phenomena such as spin-orbit coupling are accounted for by the Dirac equation:

\begin{equation}
    [c \boldsymbol{\alpha}\mathbf{p} + \beta c^{2} + V(\mathbf r) + \beta \boldsymbol{\sigma} \mathbf{B(r)}]\Psi(\mathbf r) = E\Psi(\mathbf r).
\end{equation}
The effective potential $V(\bf r)$ and the effective magnetic field $B(\mathbf r)$ are represented in this expression. The matrix $\beta$, which satisfies the condition $\beta^{2}=1$  
is of size 4x4. The $\alpha_{k}$ are 4x4 Dirac matrices which are defined by the Pauli matrices $\sigma_{k}$ ($\alpha_{k}=\sigma_{\kappa} \otimes \sigma_{\kappa}$, $k=(x,y,z)$).

Ground state computations for face-center cubic Si and Ge were conducted using the observed lattice constant  \cite{wyckoff1963crystal} of 5.431\r{A} and 5.657\AA, respectively. The local density approximation and the atomic sphere approximation were both utilized to estimate the exchange-correlation component of the potential. The resulting self-consistent potential was then used for further photoemission computations. The one-step photoemission model is utilized for the theoretical examination of the core-level photoemission process. All relevant ARPES-related aspects, such as the experimental setup and the transition matrix elements accounting for selection rules leading to the diffraction patterns under investigation, are properly addressed by the fully relativistic one-step model of photoemission. Starting from the Fermi's golden rule, the photocurrent is expressed by:
\begin{equation}
    I(\epsilon_{f},\mathbf{k_{\parallel}})=-\frac{1}{\pi} Im \int_{}^{}d\mathbf{r} \int_{}^{}d\mathbf{r^{'}} \times \Psi_{f}^{*t}(\mathbf{r}) \Delta(\mathbf{r})G_{1}(E,\mathbf{r},\mathbf{r^{'}}) \Delta^{\dagger}(\mathbf{r^{'}}) \Psi_{f}(\mathbf{r^{'}}).
\end{equation}

The initial state is described by the imaginary part of the core-level Green matrix:
\begin{equation}
    -\frac{1}{\pi} Im G_{1}(E) = \sum_{\mathcal{K}_{m}}^{} \left\lvert \mathcal{K}_{m}\right\rangle \frac{\Sigma_c}{\left( E-\varepsilon_{\mathcal{K}_{m}}\right)^{2} + \Sigma_{c}^{2}} \left\langle \mathcal{K}_{m} \right\rvert,
\end{equation}
where $\mathcal{K}_{m}$ is a relativistic core-level wave function with the energy eigenvalue $\varepsilon_{\mathcal{K}_{m}}$ and $\mathcal{K}=(\kappa_{m},\mu_{m})$. $\kappa_{m}$ and $\mu_{m}$ are relativistic indices. $\Sigma_{c}$ implies the imaginary part to the complex self-energy $\Sigma$ which can be utilized to phenomenologically explain potential damping processes of the initial state. The core states are defined by the spectral representation of the Green function, which is created using core-level wave functions obtained from the atomic-like Dirac equation  \cite{ebert1989fully}.

The dipole operator $\Delta(r)$ mediates the coupling of the high-energy final state with the low-energy initial states. In full formalism, it follows as:
\begin{equation}
    \Delta=E_{fi}\left( \mathbf{A_{0}}\nabla + \frac{\text{i}\omega}{c} \mathbf{\alpha}\mathbf{A_{0}}\right)V(\mathbf r) + E_{fi}\left( \mathbf{A_{0}}\nabla\right)\beta\sigma \mathbf{B}(r) + E_{fi}\frac{\omega}{c}\beta A_{0} \times \sigma \mathbf{B}(r),
\end{equation}
with $E_{fi}=-2\text{i}c/[(E_{f}+c^{2})^{2} - (E_{i}+c^{2})^{2}]$. $E_{i}$ and $E_{f}$ are the energy of initial and final states, respectively. The derivation of the expression involves applying commutator and anticommutator rules in a similar manner to the nonrelativistic case described in  \cite{potthoff1997one}. Herein, $\mathbf{A_{0}}$ is the spatially constant amplitude of the electromagnetic vector potential. The scalar potential $V(\mathbf r)$ and the magnetic exchange field $\mathbf{B}(r)$ are obtained by self-consistent electronic structure calculations performed within the \textit{ab-initio} framework of spin-density functional theory by use of the Vosko et al. \cite{vosko1980accurate} parameterization of the exchange and correlation potential. To account for, among others, impurity scattering a small constant imaginary value of $V_{0i}=0.01$ eV was used for the initial core state. For the final state, a constant imaginary value of $V_{0i}=1$ eV has been chosen again in a phenomenological way.

The final state is represented by a time-reversed LEED (TR-LEED) state from Pendry's model  \cite{pendry1974low,pendry1976theory,hopkinson1980calculation}:
\begin{equation}
    \Psi_{f}^{*t}(r)=\left\langle \epsilon, k_{\parallel}| G_{2}^{+} | r\right\rangle\/.
\end{equation}

This study focuses exclusively on single-site scattering when calculating the final state  \cite{braun2018correlation}. It is important to note that, even in this scenario, all aspects of kinematic diffraction are accounted for by expanding the scattering between the layers into a plane-wave basis, such as in (two-dimensional) reciprocal lattice vectors $G_{hkl}$. A typical example is our previous work on angle-resolved hard X-ray photoemission   \cite{gray2011probing,minar2013recent}. It is beneficial to formally divide the angular and radial components of the initial and final state wave functions, which are linked by the dipole operator. The photoemission intensity can be expressed in this manner:
\begin{equation}
    I(\epsilon_{f},\mathbf{k_{\parallel}}) \sim \sum_{j,n}^{} \sum_{\mathcal{K}_{1},\mathcal{K}_{5}}^{}A_{f\mathcal{K}_{1}}Z_{\mathcal{K}_{1},\mathcal{K}_{5}}^{j,n}A_{f\mathcal{K}_{5}}^{*},
\end{equation}
where $A_{f\mathcal{K}_{1}}$ refers to  the relativistic multiple-scattering coefficients. $Z_{\mathcal{K}_{1},\mathcal{K}_{5}}^{j,n}$ denotes the $Z$ matrix for an atom $j$ in layer $n$, which is constructed from the corresponding initial-state or final-state in conjunction with the dipole operator.

On the way to the sample surface, the outgoing electron wave interacts with the surrounding of emitters. To gain a qualitative understanding of diffraction patterns, it is crucial to distinguish the two diffraction regimes based on the relatively medium range of the kinetic energy of photoelectrons. In typical energies less than about $\sim$300 eV, backscattering is strong   \cite{fadley1987photoelectron,fadley1994photoelectron} and it can be utilized in various ways to gather structural data about atoms that are positioned "behind" the emitter when observed by the detector  \cite{kaduwela1991application}. When the kinetic energy is $\sim$500 eV or higher, the forward scattering (zeroth-order scattering) mode is dominant  \cite{fadley1994photoelectron}. A while back, Thompson and Fadley  \cite{thompson1984x} conducted theoretical simulations to contrast the XPD at 1 and 10 keV for the emission of C 1s from a CO molecule oriented vertically. This means that sharper forward scattering diffraction characteristics are typically anticipated at higher energies. Nevertheless, the higher-order diffraction characteristics are weaker. Due to the significant kinetic energy, forward scattering is predominant in the cases under our current discussion. This treatment is sketched via the inset in Fig.~\ref{fig:Fig2-1.jpg}b where single-site forward scattering is employed within each layer. Between layers, multiple scattering is taken into account. To obtain better visualization, we make use of surface atoms from the side view. Instead of the many scattering paths (short blue arrows) that need to be treated in low-energy calculations, the forward-scattering paths (long blue arrows) strongly prevail. In addition, two examples for different reciprocal lattice vectors $\vec{G}_{hkl}$ illustrate schematically the analog to the model of Bragg reflection at lattice planes.
Under this regime, the diffraction pattern is mainly characterized by features resembling "Kikuchi patterns", such as lines and bands typically seen in high-energy electron diffraction. In order to favorably replicate the results from our experiments, we utilized at least 137 $\vec{G}_{hkl}$ for this expansion in the calculations presented here. This approach yielded results that numerically converged with respect to the Kikuchi diffraction pattern. The angular-momentum cutoff, a second convergence parameter of our calculations, necessitates terminating orbital angular momenta at a specific value due to constraints on computing time and memory. We conducted convergence experiments up to $l_{max}=11$ and observed that for $l_{max}>3$, the positions of the photoemission peaks and, importantly for the current study, the CDAD patterns remained qualitatively unchanged. We verified this convergence test using \underline{MsSpec}  \cite{sebilleau2011msspec}, a real-space cluster approach. Undoubtedly, we are aware that we would need to approach $l_{max}>30$ (as calculated from the height of centrifugal barrier  \cite{sebilleau2018multiple}) in order to make a quantitative comparison between experimental and theoretical intensities for the high kinetic energies of electrons examined here.

\subsection{Experiment}
The experiments described below were performed using the novel full-field imaging photoemission technique called momentum microscopy (MM). Here we give a brief overview of the development of this technique and information on the used instrument; for a detailed description, see Ref. \cite{tkach2023circular}. The MM method is related to photoemission electron microscopy (PEEM), which aims at high-resolution real-space imaging \cite{bauer2012brief}. The MM development was driven by the goal of increasing the visible fraction of the k-space for ARPES measurements. MM provides high-resolution images of the electronic structure (dispersing valence bands, and core-level patterns with their diffraction fine structure) in momentum space. In the language of electron microscopy, the momentum image is nothing more than the so-called 'reciprocal image'. While normal ARPES analyzers typically capture angular ranges $<$ 10\textdegree\ (30\textdegree with wide-angle lenses), MMs can image large intervals up to the full 2$\pi$ solid angle at low kinetic energies. This is enabled by a strong electrostatic field (extractor field) that collects photoelectrons over a wide range of solid angles, forming a k-image in the backfocal plane of the objective.

The MM family includes instruments with double-hemisphere \cite{tusche2015spin}, single-hemisphere \cite{schonhense2020single}, and time-of-flight (ToF) \cite{medjanik2017direct} energy analyzers. The hemisphere-based instruments provide full-field ($k_{x}$,$k_{y}$) electron distributions, resolving $10^{4}$-$10^{5}$ data pixels. ToF instruments simultaneously capture a specific energy band, typically 4-8 eV wide, and can resolve $>$100 energy slices. This number is mainly limited by the photon bandwidth. The 3D ($k_{x}$,$k_{y}$,$E_{B}$) acquisition scheme of ToF-MMs can provide up to $10^{6}$ resolved data voxels. Momentum resolutions in the range of 0.004 \r{A}$^{-1}$, as shown in \cite{schonhense2020single}, can compete with those of conventional high-resolution ARPES analyzers \cite{tamai2013spin}. After a decade of development, the energy resolution of a current single hemisphere and ToF MMs is 4.2 and 9 meV, respectively \cite{schonhense2021time}. This is sufficient for most experiments but still does not reach the sub-meV resolution level of classical hemispherical analyzers (which are looking back on a century of development).

XPD appears in MM experiments in the form of 2D patterns rich in detail with filigree fine structure. In this case, the size of the observable energy-momentum parameter space is more important than the resolution, since it determines the number of visible Kikuchi bands. The advantage over other methods is that MMs directly observe fields of view up to $>$ 15 \r{A}$^{-1}$ and up to very high kinetic energies. As a prototypical case, graphite has been studied in the hard X-ray range from 3 to 8 keV \cite{fedchenko2019high}. As in a real space microscope, the lens optics of an MM allow zooming into details for ultimate momentum resolution. In addition, the instrument can be easily switched to real-space (PEEM) imaging to locate the probing photon spot and check the quality of the sample surface. Photoelectron diffraction in MM has been studied over a wide spectral range including the VUV, soft and hard X-rays \cite{schonhense2020momentum}. Surprisingly rich patterns have been observed at kinetic energies as low as 120 eV for Ge 2p and 3p using the hybrid MM (hemispherical analyzer with ToF booster) at the DIAMOND Light Source, Didcot, UK \cite{schmitt2024hybrid}. XPD appears as an unwanted effect in valence band mapping in the hard X-ray region. Pronounced diffraction signatures are imprinted on the ARPES patterns. Thanks to the full-field imaging technique, the modulations can be eliminated by multiplicative correction as described in \cite{babenkov2019high}.

Progress on the electron detector side towards 2D and 3D recording has been accompanied by advances in soft and hard X-ray synchrotron beamlines and X-ray free electron lasers. At the beamline P22 of PETRA III a total energy resolution (combined resolution of ToF analyzer and photon bandwidth) of 62 meV at 6 keV (resolving power $10^{5}$) was achieved \cite{medjanik2019progress}. At the free-electron laser FLASH (DESY, Hamburg) a similar ToF momentum microscope is operated \cite{kutnyakhov2020time} and the first time-resolved XPS and XPD experiments with fs resolution at soft X-ray energies have been performed \cite{curcio2021ultrafast,dendzik2020observation,pressacco2021subpicosecond}. These technical advances on both the source and detector side have led to an enormous increase in the information content since the early days when CDAD was discovered with rotating electron spectrometers \cite{westphal1989circular,schonhense1990circular}, confirming theoretical models of Ritchie \cite{ritchie1975theoretical}, Cherepkov and Mc Koy et al. \cite{cherepkov1982circular,dubs1985circular}.

The CDAD measurements in subsection \ref{comparison} were performed with the ToF-MM at beamline P22, which provides high circular polarization. This instrument is equipped with a dodecapole energy prefilter \cite{tkach2023circular}, which suppresses background electrons from higher orders of the monochromator/undulator. The photon energy was set to 6 keV, where the circular polarization is about 90\%. We used the Si(111) monochromator crystal, which provides a bandwidth of about 500 meV, being of the order of the natural linewidths of the core-level signals. With the ToF analyzer set to the core level of interest, count rates were up to several $10^{6}$ counts per second. Under these conditions, the diffraction patterns could be seen already in real time at a frame rate of 1/s. Typical exposure times were 30 minutes for each photon helicity (RCP, LCP) yielding the data arrays I\textsubscript{RCP}($k_{x},k_{y}$) and I\textsubscript{LCP}($k_{x},k_{y}$). The CDAD difference and asymmetry images are derived from the equations CDAD = I\textsubscript{RCP} - I\textsubscript{LCP} and A\textsubscript{CDAD} = (I\textsubscript{RCP}($k_{x},k_{y}$) - I\textsubscript{LCP}($k_{x},k_{y}$))/(I\textsubscript{RCP}($k_{x},k_{y}$) + I\textsubscript{LCP}($k_{x},k_{y}$)), respectively. These quantities are determined pixel-by-pixel after appropriate binning. In some cases, a mild Gaussian smoothing of the raw images was employed for noise reduction.

\section*{Data availability}
Access to the data that support this study is available from the corresponding author upon reasonable request.

\section*{Code availability}
The code utilized for this study is accessible through the corresponding author upon reasonable request.

\section*{Acknowledgements}
This work was supported by the project Quantum materials for applications in sustainable technologies (QM4ST), funded as project \texttt{No. CZ.02.01.01/00/22\_008/0004572} by Programme Johannes Amos Commenius, call Excellent Research (T.-P.V., J.M., A.P.) and the Czech Science Foundation Grant No. GA \v{C}R 23-04746S (T.-P.V.). In addition, T.-P.V, S.T., D.S., A.P., and J.M. acknowledge partial funding from Horizon Europe MSCA Doctoral network grant n.101073486, EUSpecLab, funded by the European Union. This work was also supported by the Deutsche Forschungsgemeinschaft, Grant No. TRR288–422213477 (Project B04), and by the Federal Ministry of Education and Research (BMBF, Project 05K22UM2). A.W. was supported by the Polish National Science Centre (NCN), grant no. 2020/37/B/ST5/03669.

\section*{Author contributions}
PED reformulations and calculations in the SPRKKR package were performed by T.-P.V. and J.M. Convergence calculations using the \underline{MsSpec} code were done by T.-P.V with the assistance of S.T. and D.S. T.-P.V. and J.M. conducted the data analysis of the theoretical and experimental PED data with the support from J.B., A.P., G.S., H.-J.E., O.T. and A.W. The experiments were done by O.T., O.F., Y.L., D.V., H.-J.E., and G.S. T.-P.V. wrote the original draft. All authors contributed to the writing, review, and editing of the manuscript.

\section*{Competing interests}
The authors declare no competing interests.


\bibliography{Reference.bib}

\afterpage{%
\begin{figure}[H]
\caption{The SPRKKR convergence test for Si 2p$_{3/2}$.}
\centering
\includegraphics[scale=0.31]{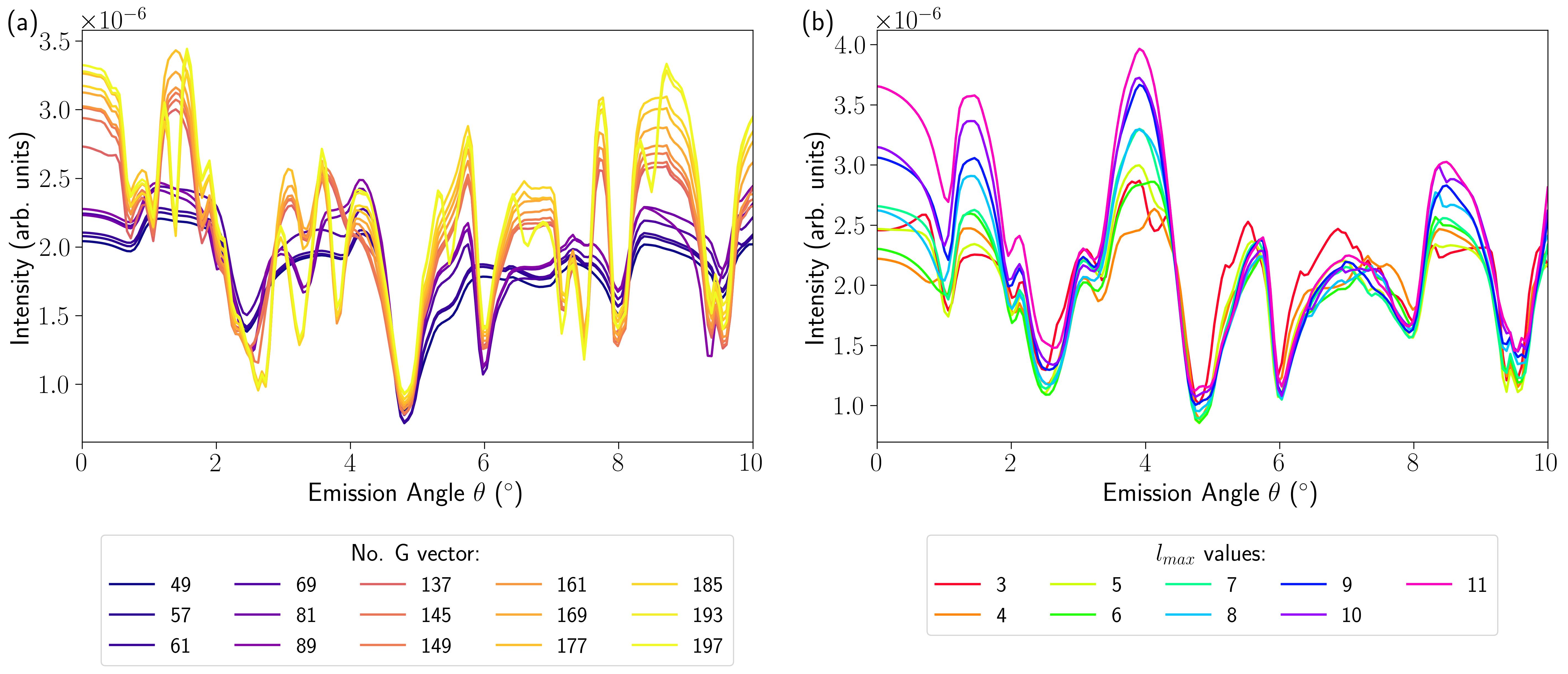}
\caption*{(a) The intensity computed as a function of emission angles with different numbers of $\vec{G}_{hkl}$ (from 49 to 197) with $l_{max}=4$. (b) The intensity computed as a function of emission angles with different numbers of $l_{max}$  values (from 3 to 11) with 101 $\vec{G}_{hkl}$.}
\label{fig:Fig3-1-1.jpg}
\end{figure}
\clearpage}

\afterpage{%
\begin{figure}[H]
\caption{Calculated total-intensity patterns as a function of $\vec{G}_{hkl}$.}
\centering
\includegraphics[scale=0.4]{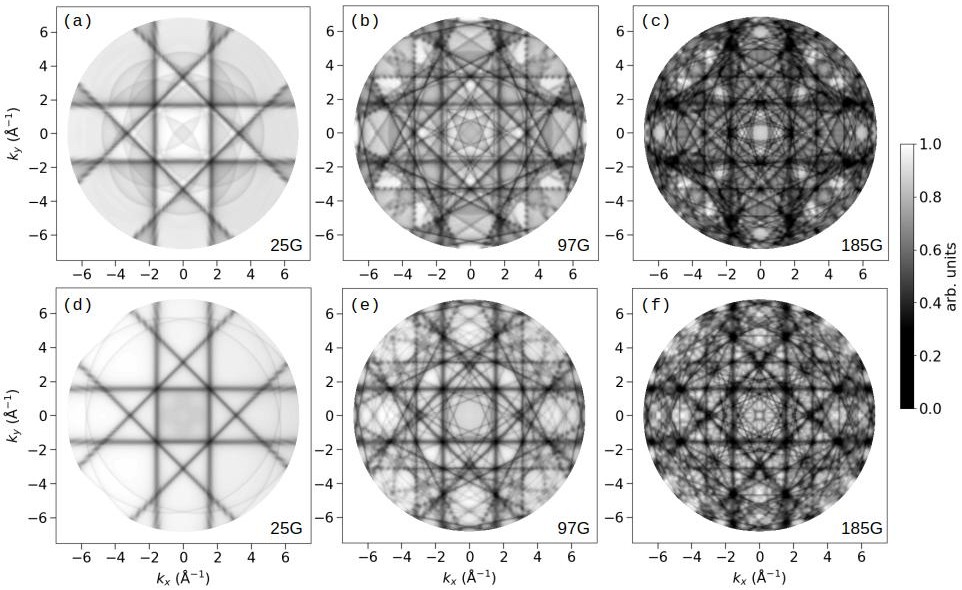}
\caption*{(a-c) Total intensity for Si 2p\textsubscript{3/2}. (d-f) Total intensity for Ge 3p\textsubscript{3/2}. The different numbers of $\vec{G}_{hkl}$ are mentioned in the panel. Emission at the final-state energy $E_{Final}=5880$ eV and photon energy $h\nu=6000$ eV with $l_{max}=4$.}
\label{fig:Fig3-1-3.jpg}
\end{figure}
\clearpage}

\afterpage{%
\begin{figure}[H]
\caption{Calculated total-intensity patterns as a function of $l_{max}$.}
\centering
\includegraphics[scale=0.4]{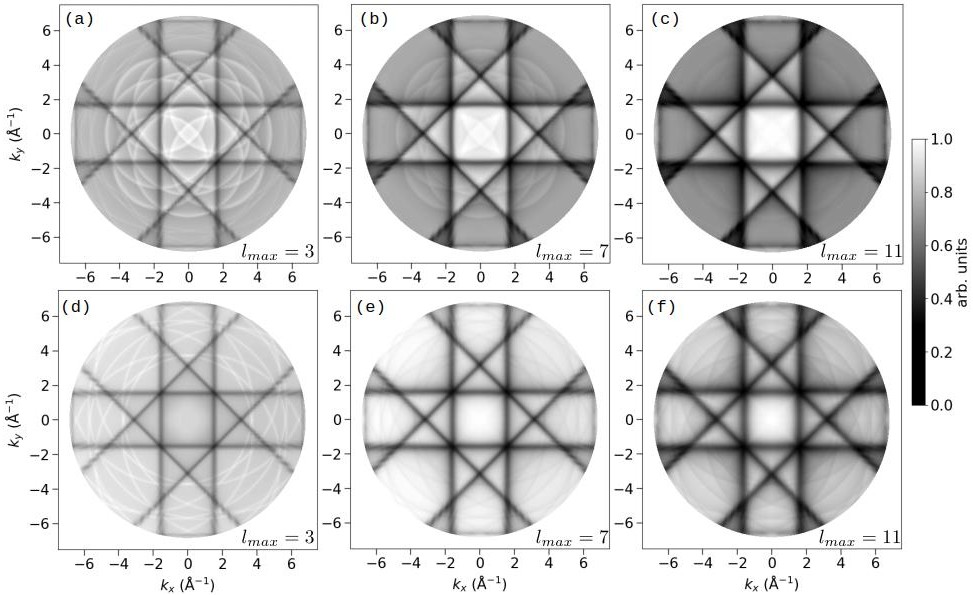}
\caption*{(a-c) Total intensity for Si 2p\textsubscript{3/2}. (d-f) Total intensity for Ge 3p\textsubscript{3/2}.  The different numbers of $l_{max}$ are mentioned in the panel. Emission at the final-state energy $E_{Final}=5880$ eV and photon energy $h\nu=6000$ eV with 37 $\vec{G}_{hkl}$.}
\label{fig:Fig3-1-4.jpg}
\end{figure}
\clearpage}

\afterpage{%
\begin{figure}[H]
\caption{The convergence test done by the \underline{MsSpec} package.}
\centering
\includegraphics[scale=0.41]{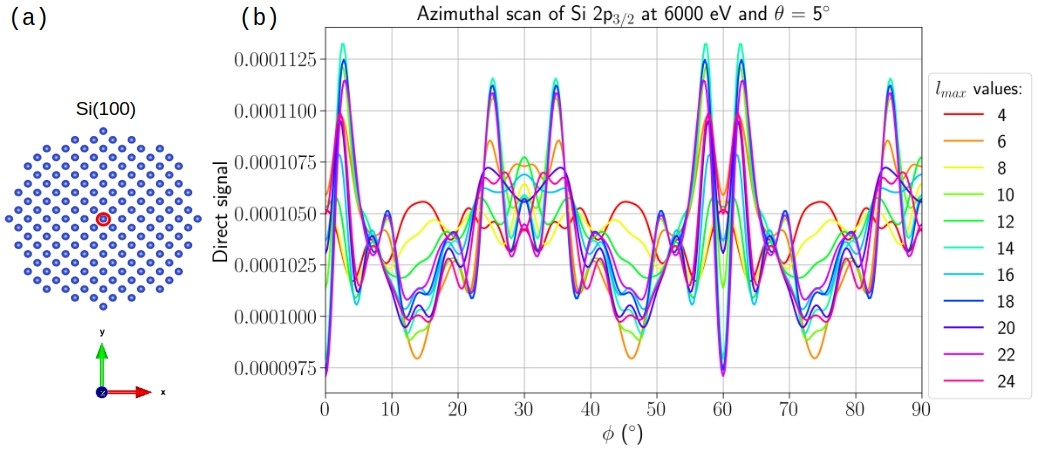}
\caption*{(a) A cluster model of Si (100). The red circle depicts the emitter. (b) The cross-section calculated as a function of $l_{max}$ values for Si 2p$_{3/2}$.}
\label{fig:Fig3-1-2.jpg}
\end{figure}
\clearpage}

\afterpage{%
\begin{figure}[H]
\caption{Sequence of calculated total intensity as a function of final-state energies.}
\centering
\includegraphics[scale=0.41]{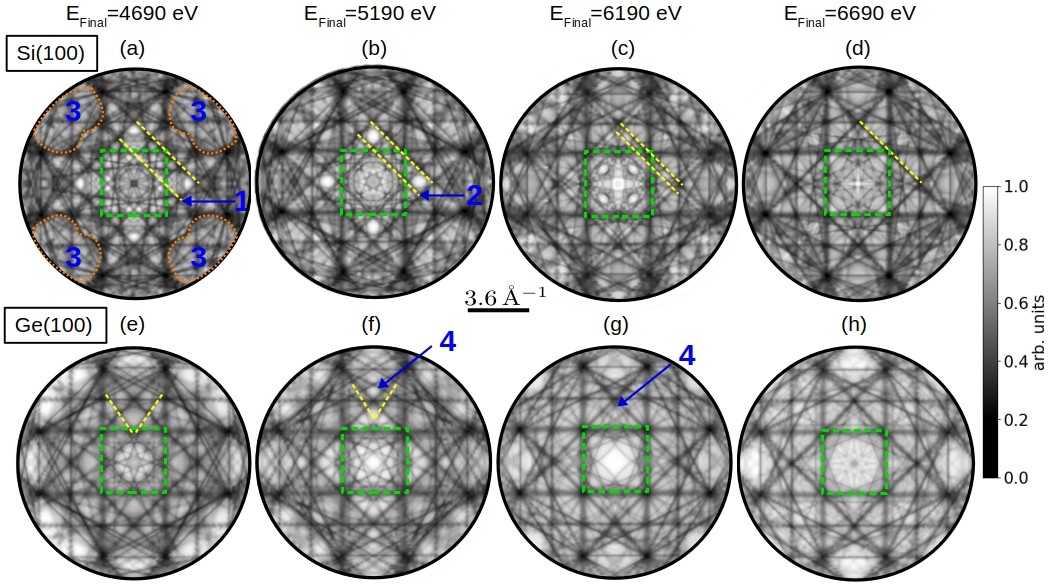}
\vspace*{-4mm}
\caption*{(a-d) The diffractogram of Si 2p\textsubscript{3/2}. (e-h) The diffractogram of Ge $3p_{3/2}$. Calculations are performed with 161 $\vec{G}_{hkl}$ for Si and 113 $\vec{G}_{hkl}$ for Ge at $l_{max}=4$ and $h\nu$= 6 keV. The final-state energy $E_{Final}$ is between 4.69 and 6.69 keV. All plots are made at the same color scale. Green dash squares indicate the central zone of studied structures. The zone axis is [100], directly out of the page.}
\label{fig:Fig3-2-1.jpg}
\end{figure}
\clearpage}

\afterpage{%
\begin{figure}[H]
\caption{Comparison between measured and calculated patterns of Si 2p$_{3/2}$ at $h\nu$= 6 keV.}
\centering
\includegraphics[scale=0.36]{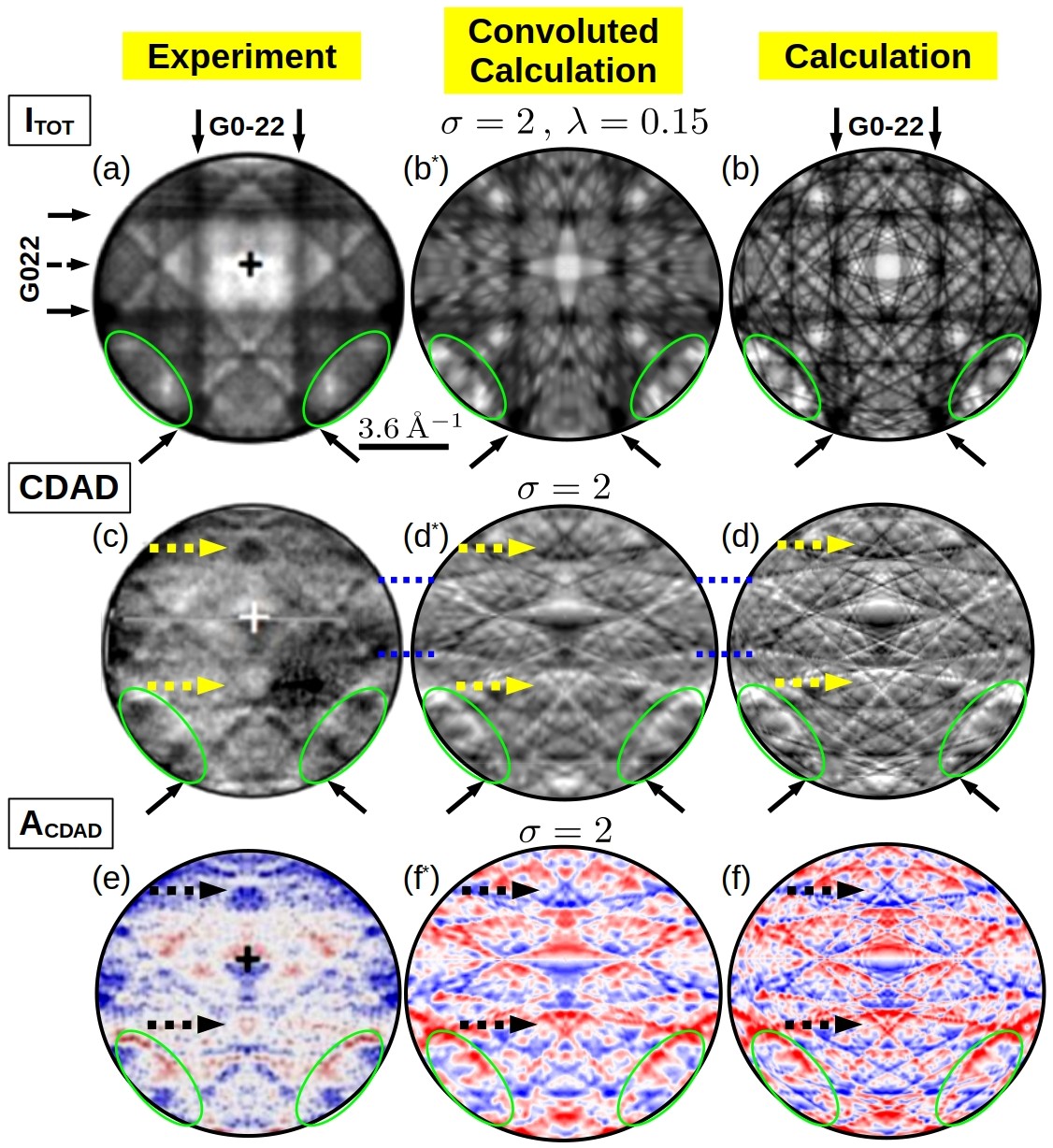}
\caption*{The top, middle, and bottom rows indicate the total intensity I\textsubscript{TOT}, intensity difference CDAD and CDAD asymmetry A\textsubscript{CDAD}. From left to right, there are measured (a, c, e), convoluted-calculated (\text{b$^{*}$}, \text{d$^{*}$}, \text{f$^{*}$}) and purely calculated patterns (b, d, f). Computational results are performed at $l_{max}=4$ with 193 $\vec{G}_{hkl}$. To smooth the computed data, convolution is applied by the Gaussian filter with the standard deviation ($\sigma$) and the Perona-Malik filter (anisotropic diffusion) with the "weight" parameter ($\lambda$).}
\label{fig:Fig3-4-1.jpg}
\end{figure}
\clearpage}

\afterpage{%
\begin{figure}[H]
\caption{Comparison between measured and calculated patterns of Si 2p$_{3/2}$ at $h\nu$= 6 keV.}
\centering
\includegraphics[scale=0.6]{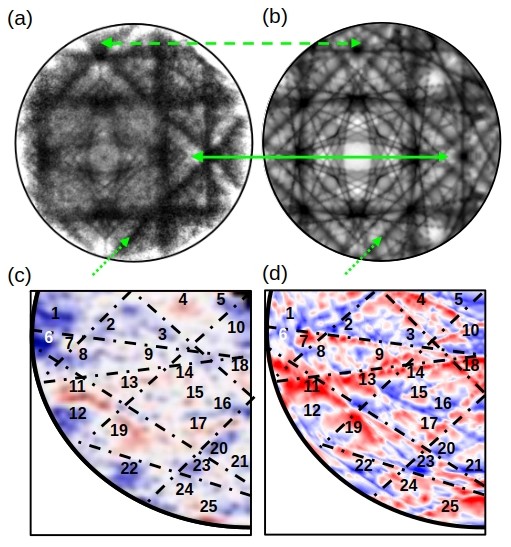}
\caption*{(a) Measured total intensity pattern I\textsubscript{TOT} by higher resolution, compared to Fig.~\ref{fig:Fig3-4-1.jpg}a. (b) Total intensity calculation by OSM. (c,d) CDAD asymmetry displayed by the quadrant of Fig.~\ref{fig:Fig3-4-1.jpg} (e,f). Computational results are performed with 193 $\vec{G}_{hkl}$.}
\label{fig:Fig3-4-1-1.jpg}
\end{figure}
\clearpage}

\afterpage{%
\begin{figure}[H]
\caption{Comparison between measured and calculated patterns of Ge 2p$_{3/2}$ at $h\nu$= 6 keV.}
\centering
\includegraphics[scale=0.35]{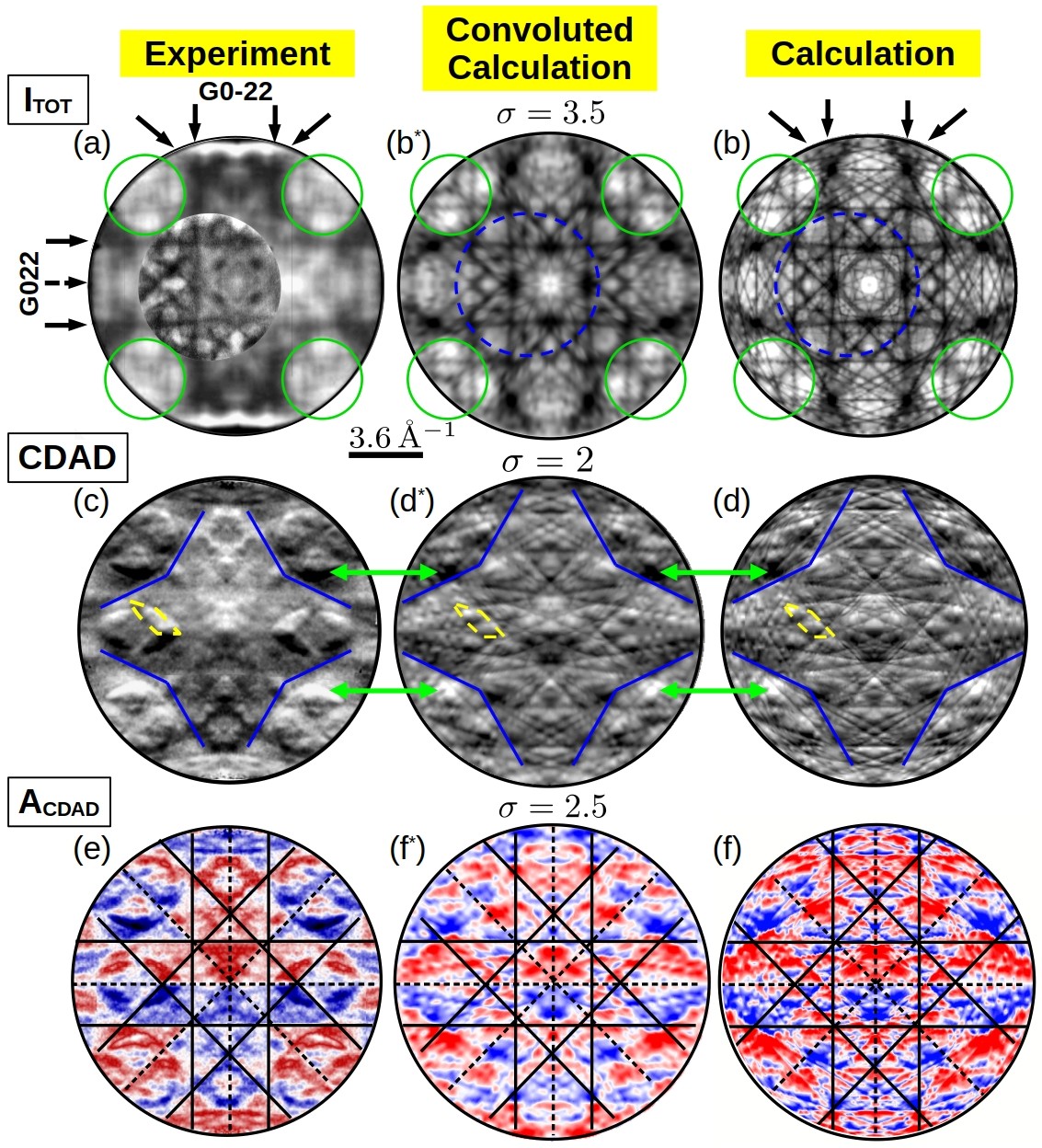}
\caption*{The top, middle, and bottom rows indicate the total intensity I\textsubscript{TOT}, intensity difference CDAD and CDAD asymmetry A\textsubscript{CDAD}. From left to right, there are measured (a, c, e), convoluted-calculated (\text{b$^{*}$}, \text{d$^{*}$}, \text{f$^{*}$}) and purely calculated patterns (b, d, f). Computational results are performed with 177 $\vec{G}_{hkl}$. To smooth the computed data, convolution is applied by the Gaussian filter with the standard deviation $\sigma$.}
\label{fig:Fig3-4-2.jpg}
\end{figure}
\clearpage}

\afterpage{%
\begin{figure}[H]
\caption{Comparison between measured and calculated patterns of Ge 3p$_{3/2}$ at $h\nu$= 6 keV.}
\centering
\includegraphics[scale=0.49]{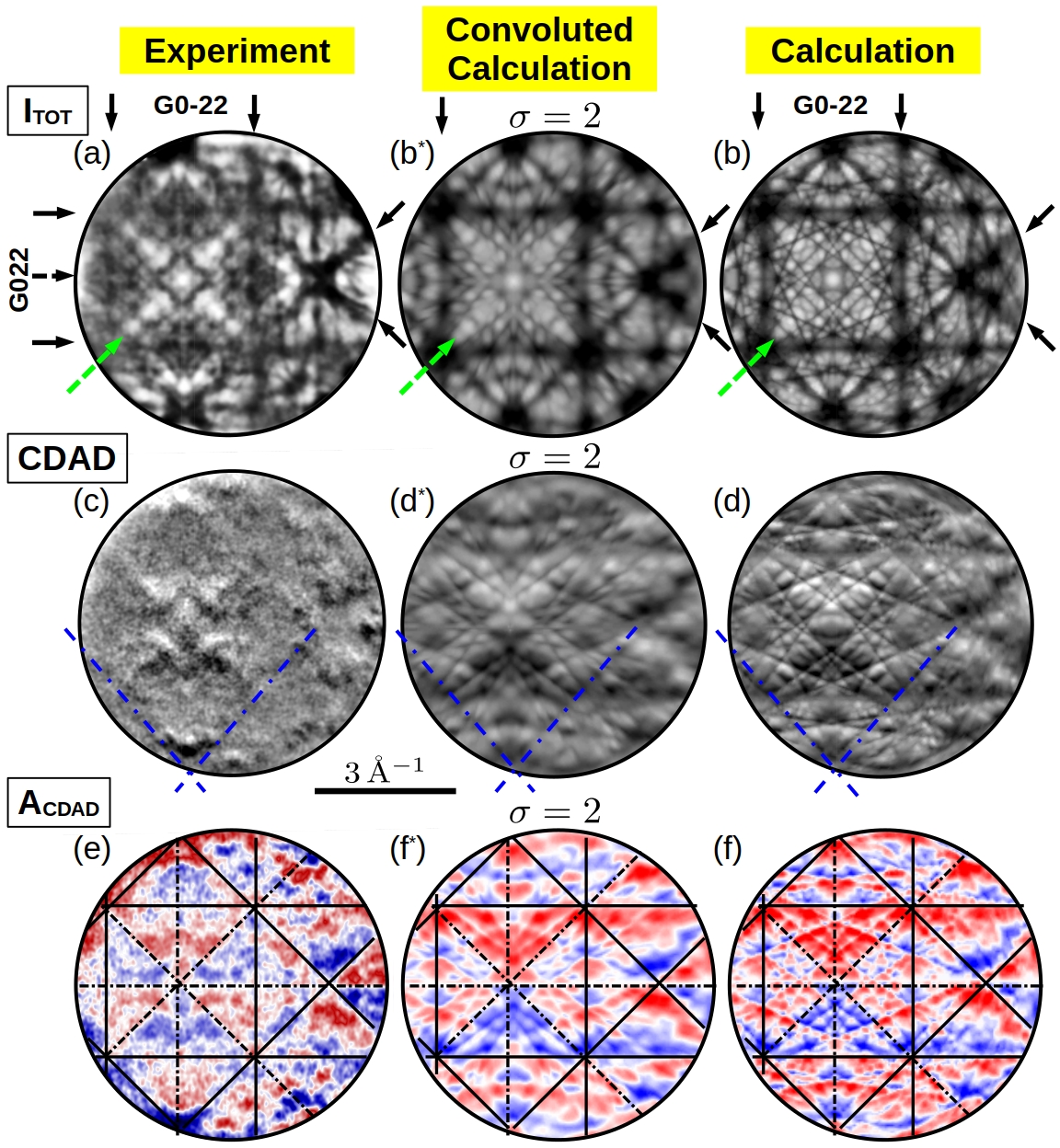}
\caption*{The top, middle, and bottom rows indicate the total intensity I\textsubscript{TOT}, intensity difference CDAD and CDAD asymmetry A\textsubscript{CDAD}. From left to right, there are measured (a, c, e), convoluted-calculated (\text{b$^{*}$}, \text{d$^{*}$}, \text{f$^{*}$}) and purely calculated patterns (b, d, f). Computational results are performed with 241 $\vec{G}_{hkl}$. To smooth the computed data, convolution is applied by the Gaussian filter with the standard deviation sigma ($\sigma$).}
\label{fig:Fig3-4-2-1.jpg}
\end{figure}
\clearpage}

\afterpage{%
\begin{figure}[H]
\caption{Schematic representation of PED computational methods.}
\centering
\includegraphics[scale=0.41]{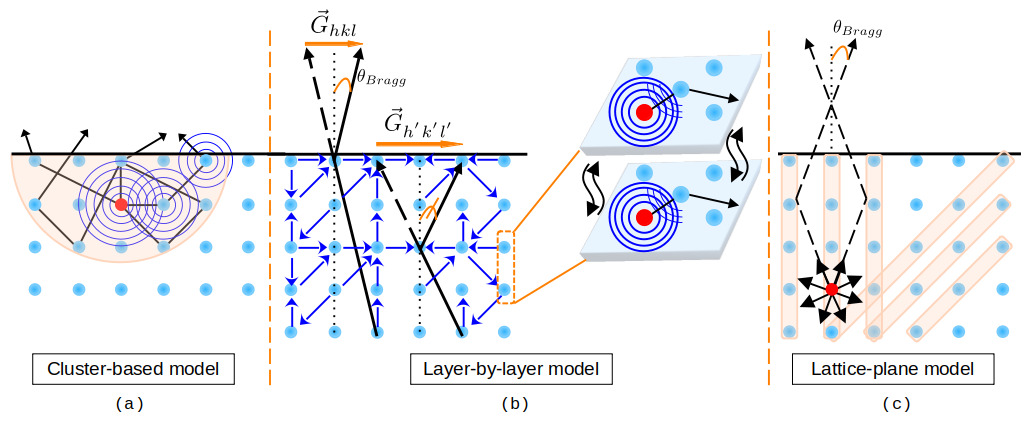}
\caption*{(a) The cluster-based model. (b) The layer-by-layer model. (c) The lattice-plane model. Atoms are represented by blue. Localized sources are indicated by red and surrounded by spherical waves. Black/blue arrows illustrate single-site/multiple scattering paths. } 
\label{fig:Fig2-1.jpg}
\end{figure}
\clearpage}
\end{document}